\documentclass[manuscript]{aastex}

\usepackage{txfonts}
\usepackage{longtable}
\usepackage{rotating}
\usepackage{natbib}
\usepackage{graphicx}
\usepackage{graphics}
\usepackage{psfrag}
\usepackage{amssymb}
\bibpunct{(}{)}{;}{a}{}{,}

\def\kms{$\mathrm{km\,s}^{-1}$}

\def\llm{{\sc LLmodels}}

\def\cs{$\mathrm{count\,s}^{-1}$}

\shorttitle{Metals in the exosphere of WASP-12b}
\shortauthors{Fossati et al.}

\begin{document}

\title{Metals in the exosphere of the highly-irradiated planet
WASP-12b$^1$}
\altaffiltext{1}{Based on observations made with the NASA/ESA Hubble Space
Telescope, obtained MAST at the Space Telescope Science Institute, which is
operated by the Association of Universities for Research in Astronomy, Inc.,
under NASA contract NAS 5-26555. These observations are associated with
program \#11651.}

\author{L. Fossati and C.A. Haswell}
\affil{Department of Physics and Astronomy, Open University,
	Walton Hall, Milton Keynes MK7 6AA, UK}
\email{l.fossati@open.ac.uk,C.A.Haswell@open.ac.uk}
\and
\author{C. S. Froning\altaffilmark{2}}
\affil{Center for Astrophysics and Space Astronomy,
	University of Colorado, 593 UCB, Boulder, CO 80309-0593, USA}
\email{cynthia.froning@colorado.edu}
\and
\author{L. Hebb}
\affil{Department of Physics and Astronomy, Vanderbilt University, 
	6301 Stevenson Center Nashville, TN 37235, USA}
\email{leslie.hebb@vanderbilt.edu}
\and
\author{S. Holmes and U. Kolb}
\affil{Department of Physics and Astronomy, Open University,
	Walton Hall, Milton Keynes MK7 6AA, UK}
\email{s.holmes@open.ac.uk,U.C.Kolb@open.ac.uk}
\and
\author{Ch. Helling}
\affil{SUPA, School of Physics and Astronomy, University of St Andrews, 
	North Haugh, St Andrews KY16 9SS, UK}
\email{Christiane.Helling@st-andrews.ac.uk}
\and
\author{A. Carter}
\affil{Department of Physics and Astronomy, Open University,
	Walton Hall, Milton Keynes MK7 6AA, UK}
\email{A.Carter@open.ac.uk}
\and
\author{P. Wheatley}
\affil{Department of Physics, University of Warwick, Coventry CV4 7AL, UK}
\email{p.j.wheatley@warwick.ac.uk }
\and
\author{A.~C. Cameron}
\affil{SUPA, School of Physics and Astronomy, University of St Andrews, 
	North Haugh, St Andrews KY16 9SS, UK}
\email{acc4@st-andrews.ac.uk}
\and
\author{B. Loeillet}
\affil{Laboratoire d'Astrophysique de Marseille, BP 8, 13376 Marseille 
	Cedex 12; Universit{\'e} de Provence, CNRS (UMR 6110) and CNES, France}
\email{benoit.loeillet@oamp.fr}
\and
\author{D. Pollacco}
\affil{Astrophysics Research Centre, School of Mathematics \& Physics, 
	Queen's University, University Road, Belfast BT7 1NN, UK}
\email{d.pollacco@qub.ac.uk}
\and
\author{R. Street}
\affil{Las Cumbres Observatory, 6740 Cortona Dr. Suite 102, 
	Santa Barbara, CA 93117, USA}
\email{rstreet@lcogt.net}
\and
\author{H.~C. Stempels\altaffilmark{3}}
\affil{SUPA, School of Physics and Astronomy, University of St Andrews, 
	North Haugh, St Andrews KY16 9SS, UK}
\email{Eric.Stempels@fysast.uu.se}
\and
\author{E. Simpson}
\affil{Astrophysics Research Centre, School of Mathematics \& Physics, 
	Queen's University, University Road, Belfast BT7 1NN, UK}
\email{esimpson05@qub.ac.uk}
\and
\author{S. Udry}
\affil{Observatoire de Gen{\`e}ve, Universit{\'e} de Gen{\`e}ve, 
	51 Ch. des Maillettes, 1290 Sauverny, Switzerland}
\email{Stephane.Udry@unige.ch}
\and
\author{Y.~C. Joshi}
\affil{Astrophysics Research Centre, School of Mathematics \& Physics, 
	Queen's University, University Road, Belfast BT7 1NN, UK}
\email{y.joshi@qub.ac.uk}
\and
\author{R.~G. West}
\affil{Department of Physics \& Astronomy, University of Leicester, 
	Leicester, LE1 7RH, UK}
\email{rgw@astro.le.ac.uk}
\and
\author{I. Skillen}
\affil{Isaac Newton Group of Telescopes, Apartado de Correos 321, 
	38700 Santa Cruz de la Palma, Tenerife, Spain}
\email{wji@ing.iac.es}
\and
\author{D. Wilson\altaffilmark{4}}
\affil{Astrophysics Group, Keele University, Staffordshire, ST5 5BG, UK}
\email{david@davidmwilson.net}

\altaffiltext{2}{Department of Astrophysical and Planetary Sciences,
University of Colorado at Boulder, US}
\altaffiltext{3}{Department of Physics and Astronomy, Box 516, SE-751 20,
Uppsala, Sweden}
\altaffiltext{4}{Centre for Astrophysics \& Planetary Science, 
University of Kent, Canterbury, Kent, CT2 7NH, UK}

\begin{abstract}
We present near-UV transmission spectroscopy of the highly irradiated 
transiting exoplanet WASP-12b, obtained with the Cosmic Origins 
Spectrograph (COS) on the Hubble Space Telescope (HST). The spectra cover 
three distinct wavelength ranges: NUVA (2539--2580\,\AA); 
NUVB (2655--2696\,\AA); and NUVC (2770--2811\,\AA). Three independent 
methods all reveal enhanced transit depths attributable to
absorption by resonance lines of metals in the exosphere of WASP-12b. Light
curves of total counts in the NUVA and NUVC wavelength ranges show a 
detection at a 2.5$\sigma$ level. We detect extra absorption in the 
\ion{Mg}{2} $\lambda\lambda$2800 resonance line cores at the 2.8$\sigma$ 
level. The NUVA, NUVB and NUVC light curves imply effective radii of 
2.69$\pm$0.24\,R$_J$, 2.18$\pm$0.18\,R$_J$, and 2.66$\pm$0.22\,R$_J$ 
respectively, suggesting the planet is surrounded by an absorbing cloud 
which overfills the Roche lobe. We detect enhanced transit depths at the 
wavelengths of resonance lines of neutral sodium, tin and manganese, and 
at singly ionised ytterbium, scandium, manganese, aluminum, vanadium and 
magnesium. We also find the statistically expected number of anomalous 
transit depths at wavelengths not associated with any known resonance line. 
Our data are limited by photon noise, but taken as a whole the results are 
strong evidence for an extended absorbing exosphere surrounding the planet. 
The NUVA data exhibits an early ingress, contrary to model expectations; 
we speculate this could be due to the presence of a disk of previously 
stripped material.
\end{abstract}


\section{Introduction}
Observations of the transiting extrasolar planets HD209458b and HD189733b
revealed an enhanced transit depth at the wavelengths of several UV resonance 
lines \citep{vidal03,vidal04,lecavelier10}. These UV lines from the
ground state are sensitive probes of the presence of atomic and ionic
species. Their presence enhanced the effective radius of the planet
during transit, implying the planet is surrounded by an extended cloud
of size comparable to or larger than its Roche lobe
\citep{vidal03,vidal04,jaffel07,vidal08}. This was attributed to a
hydrodynamic `blow-off' of the planet's outer atmosphere caused by the
intense irradiation suffered by this hot Jupiter exoplanet. An alternative
explanation in which the planet is surrounded by a cloud of energetic
neutral atoms caused by interactions with the host star's stellar wind
has, however, been suggested \citep{ena08,ena10}. WASP-12b is one of the
hottest and most irradiated transiting exoplanets and orbits extremely close
to a late F-type host star \citep{hebb2009}. WASP-12b is, therefore, an
attractive target to explore the properties of the phenomenon observed 
in HD209458b, and might yield evidence distinguishing between the 
suggested underlying causes.

The initial UV observations of HD209458b were in the far UV around the
Ly\,$\alpha$ emission line. The abundance of hydrogen makes this an attractive
line to observe, but the temporal and spatial variability of stellar 
Ly\,$\alpha$ emission is a highly undesirable complicating factor.
For this reason, and to obtain better signal to noise, we observed WASP-12 in
the near-UV where there are many other resonance lines
\citep{morton1991,morton2000}, including the very strong \ion{Mg}{2} UV
resonance lines. This work became possible with the installation of the
Cosmic Origins Spectrograph (COS) on the Hubble Space
Telescope (HST) reinstating and enhancing our capabilities for UV spectroscopy. 
\section{Observations and data reduction}
The planet-hosting star WASP-12 was observed for five consecutive HST orbits 
on  2009 September 24th and 25th with COS; see \citet{green2003}, 
Green et al. (2010, in preparation) and Osterman et al. (2010, in preparation) 
for details of COS. We used the NUV G285M grating at the 2676\,\AA\ setting,
which provides non-contiguous spectra over three wavelength ranges
(NUVA: 2539--2580\,\AA, NUVB: 2655--2696\,\AA, and
NUVC: 2770--2811\,\AA) at a spectral resolution of $R\sim$20\,000, in 
TIME-TAG mode. The exposure time was 2334\,sec in the first 
HST orbit and about 3000\,sec per subsequent HST orbit. The optical 
ephemeris gives ingress during the second HST orbit and egress in the 
fourth HST orbit.

We downloaded data from MAST\footnote{\tt http://archive.stsci.edu/} adopting 
CALCOS V.2.11b\footnote{See the COS Data Handbook for more information on 
CALCOS: \tt http://www.stsci.edu/hst/cos/documents/handbooks/\\datahandbook/COS\_longdhbcover.html}
for calibration. Despite the early date of our observations, the CALCOS
reference files used were at a fairly mature stage for the NUV data. In
particular, the flat field had been updated to its flight version.
In our time series analysis we used the count rates obtained after  
background subtraction, rather than the flux calibrated spectra. The high 
quality flat-field and the relatively low background of the NUV channel, 
mean the uncertainties are dominated by 
poisson statistics. The count rates summed over wavelength are roughly
10\,\cs; 28\,\cs; and 13\,\cs\ respectively for the NUVA, NUVB, and NUVC 
ranges. The resulting signal to noise ratio (SNR) per pixel in the NUVB 
spectrum is $\sim$10 for each 3000\,sec exposure.

\begin{figure}
\includegraphics[width=90mm]{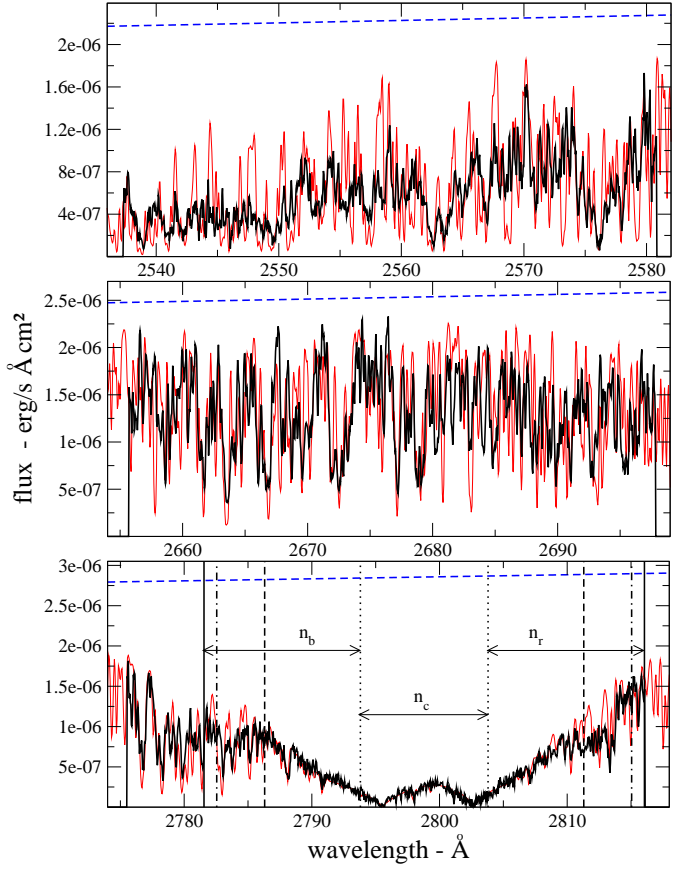}
\caption{\label{spectra} Comparison between the observed mean spectrum 
of WASP-12 (thick black line) and \llm\ synthetic fluxes (thin red line). 
The blue dashed line shows the modeled level of the stellar continuum flux.
The three observed spectral ranges are defined as NUVA, NUVB and NUVC from
top to bottom. In the bottom panel the vertical lines show the limits applied
for the wavelength regions, in laboratory wavelengths, accounted to 
produce the photometric indexes described in Sect.~\ref{charb}, showing as 
example the wavelength regions of n$_b$, n$_c$, and n$_r$.}
\end{figure}

Figure~\ref{spectra} shows the total summed spectrum in comparison with 
synthetic fluxes from the \llm\ stellar model atmosphere 
code \citep{llm}, assuming the fundamental parameters and metallicity 
given by \citet{hebb2009}.
We used the VALD database \citep{vald1,vald2,vald3} for atomic line 
parameters and SYNTH3 \citep{synth3} for spectral synthesis.
All three regions are strongly affected by many blended photospheric 
absorption lines; we observe no unabsorbed stellar continuum. The NUVB 
region is closest to the continuum, while the NUVC region is strongly 
absorbed by the \ion{Mg}{2} doublet at 2795.5\,\AA\ and 2802.7\,\AA.
\clearpage
\section{Detection of a wavelength dependent planet transit}
We expect the planet's atmosphere to absorb particularly in the resonance 
lines of abundant elements. We used three methods to examine the data for 
wavelength-dependence of the transit light curve. 
\subsection{The \ion{Mg}{2} lines}\label{charb}
The most prominent observed lines in the stellar photosphere are the 
\ion{Mg}{2} lines, and we might expect these strong lines to be detectable 
in the planet's atmosphere too. We adopted the method pioneered by 
\citet{charbonneau2002} in their detection of the sodium D lines in the 
atmosphere of HD209458b. We divided the NUVC data, which is centred on 
the \ion{Mg}{2} resonance lines, into ``blue" (b), ``red" (r),
and ``center" (c) spectral regions. We tried three different widths of the 
center band, ``narrow" (n), ``medium" (m), and ``wide" (w); see 
Fig.~\ref{spectra} and Table~\ref{table_bands}.
For each of these bands we produced a photometric time series, and the 
associated uncertainty based on Poisson statistics. Each photometric index 
was obtained by summing the observed counts over the given wavelength range. 
In this way ``n$_b(t)$" indicates the count rate in the blue side ``narrow'' 
set at the time $t$.
\clearpage
\begin{table}
\caption{\label{table_bands} Limits adopted to define the analysed wavelength
regions around the \ion{Mg}{2} resonance lines.}
\begin{tabular}{lclclc}
\tableline\tableline
Band & Wavelength  & Band & Wavelength  & Band & Wavelength  \\
     & range [\AA] &      & range [\AA] &      & range [\AA] \\
\tableline
$n_{b}$ & 2782.75 - 2795    & $m_{b}$ & 2782.75 - 2787.5  & $w_{b}$ & 2782.75 - 2783.75 \\
$n_{r}$ & 2805    - 2817.25 & $m_{r}$ & 2812.5  - 2817.25 & $w_{r}$ & 2816.25 - 2817.25 \\
$n_{c}$ & 2795    - 2805    & $m_{c}$ & 2787.5  - 2812.5  & $w_{c}$ & 2783.75 - 2816.25 \\
\tableline
\end{tabular}
\end{table}

The stellar limb darkening could potentially cause a color-dependent transit 
shape \citep[e.g,][]{brown2001}. To assess this we calculated the
difference of the blue and red spectral regions for the ``n", ``m", and ``w"
bands as a function of time \citep[see Eq.~1 of][]{charbonneau2002}.
We looked for variations in the transit depth due to the stellar limb 
darkening calculating the difference between the mean photometric indexes 
obtained in- and out-of-transit \citep[see Eq.~2 of][]{charbonneau2002}. 
All values we obtained were clearly consistent with no variation.

To examine time dependence using \citet{charbonneau2002}'s method
we calculated in each band (``n", ``m", and ``w") the difference between the 
light curve of the central band and the mean light curve of the blue and red 
bands:
\begin{equation}
\begin{array}{ccccccccc}
n_{Mg}(t) & = & n_c(t) & - & [n_b(t) & + & n_r(t)] & / & 2\phm{.}\\
m_{Mg}(t) & = & m_c(t) & - & [m_b(t) & + & m_r(t)] & / & 2\phm{.}\\
w_{Mg}(t) & = & w_c(t) & - & [w_b(t) & + & w_r(t)] & / & 2.
\end{array}
\end{equation}
In this way, we removed any limb darkening dependence. Again, the time 
series have RMS scatter consistent with 
photon noise:
($\sigma[n_{Mg}(t_{out})]\,\sim\,\sigma[m_{Mg}(t_{out})]\,\sim\,
\sigma[w_{Mg}(t_{out})]\,\sim\,3.4\times10^{-3}$\,\cs).
We then calculated the difference between the mean in-transit and
out-of-transit flux: 
\begin{equation}\label{mg2depth}
\begin{array}{ccccccc}
\Delta n_{Mg} & = & \overline{n_{Mg}(t_{in})} & - & \overline{n_{Mg}(t_{out})} & = & (3.5 \pm 4.1) \times 10^{-3}\,\mathrm{count\,s}^{-1}\phm{.}\\
\Delta m_{Mg} & = & \overline{m_{Mg}(t_{in})} & - & \overline{m_{Mg}(t_{out})} & = & (-4.7 \pm 4.1) \times 10^{-3}\,\mathrm{count\,s}^{-1}\phm{.}\\
\Delta w_{Mg} & = & \overline{w_{Mg}(t_{in})} & - & \overline{w_{Mg}(t_{out})} & = & (-11.4 \pm 4.1) \times 10^{-3}\,\mathrm{count\,s}^{-1}.
\end{array}
\end{equation}
These results show the detection of a deeper transit in the ``m" and ``w" 
bands at 1.1$\sigma$ and 2.8$\sigma$, respectively. Since the value obtained 
in the ``n" band is comparable to the resulting photon noise error bar we 
believe that the non-detection is due to the very low signal level in $n_c$
along with absorption occurring in the wide $n_r$ and $n_b$ bands. The size 
and the significance of the detection increases as the signal 
included in the center band increases, just as we would expect if the 
enhanced transit depth in the \ion{Mg}{2} doublet is genuine.
\subsection{The transit light curve}\label{sec_lightcurve}
\begin{figure}
\includegraphics[width=13cm]{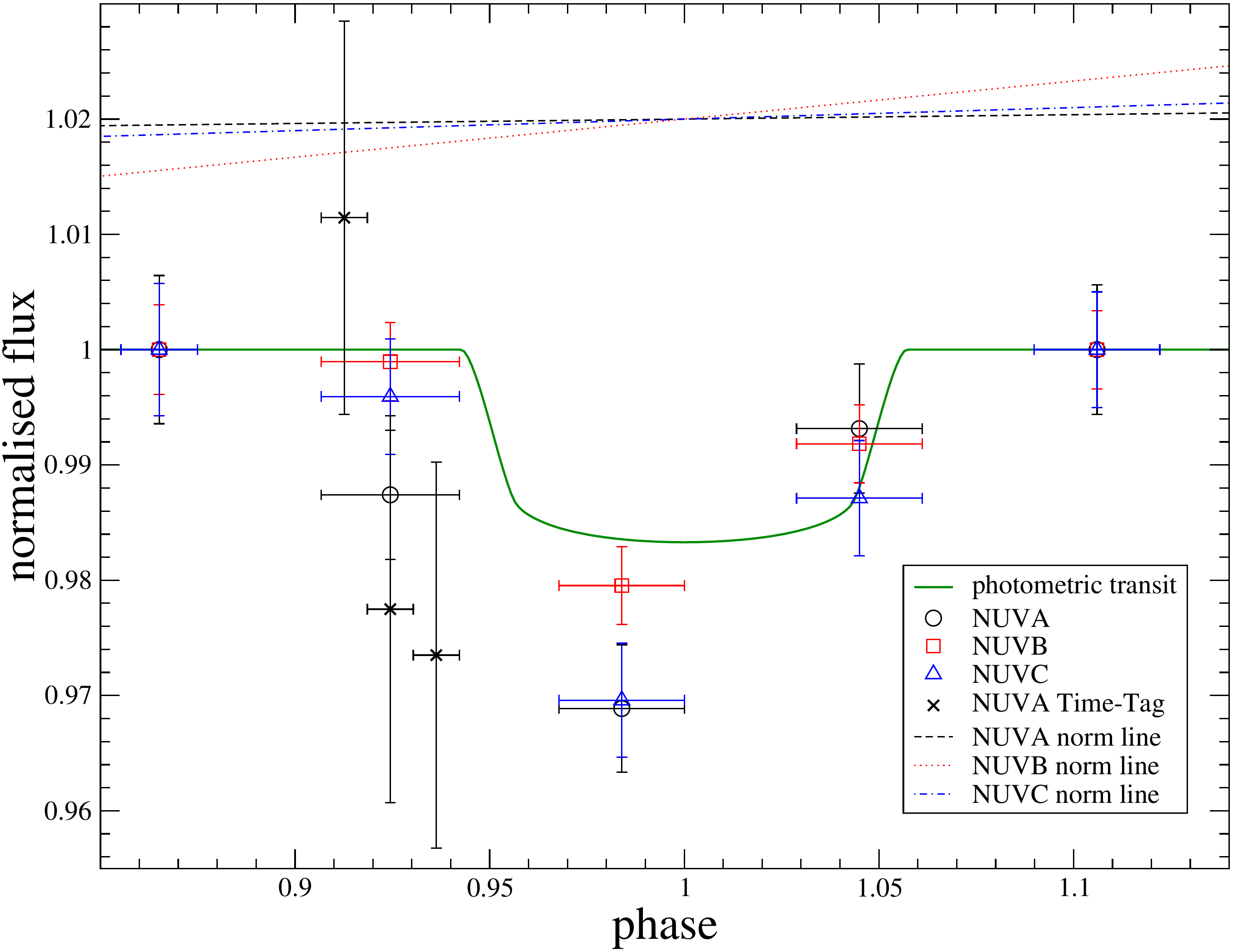}
\caption{\label{lightcurve} Light curve obtained for each observed wavelength
range (NUVA: open black circles - NUVB: open red squares - NUVC: open blue
triangles). The horizontal error bar defines the orbital phase range covered by
each observation. The vertical uncertainty comes from a Poissonian treatment of
the error bars. The full green line shows the MCMC fit to the optical transit 
light curve \citep{hebb2009}. The black crosses show the NUVA
spectral range split into three equally exposed sub-exposures. 
Lines indicate the normalisation gradient applied.}
\end{figure}

We compared the light curves obtained for each observed wavelength range and 
the one calculated from visible photometry, as shown in Fig.~\ref{lightcurve}.
\clearpage
The NUVB wavelength range is the closest to the continuum and shows a transit
depth that matches, at $\sim1\sigma$, the transit light curve derived by
\citet{hebb2009} from optical photometry. In the NUVA and NUVC wavelength
ranges we obtained a deeper transit at about 2.5$\sigma$ level.
These three light curves were normalised to the line passing through the
out-of-transit photometric points (first and fifth exposures).
The slope of the three normalisation lines are $3.8\times10^{-3}$ for the 
NUVA region, $3.3\times10^{-2}$ for the NUVB region, and $1.0\times10^{-2}$ 
for the NUVC region. These values are small enough that the applied 
normalisation did not change the transit shape.

The NUVC spectral region is clearly dominated by the \ion{Mg}{2} resonance
lines that are likely to be  responsible for the detected extra depth in the
transit light curve. The NUVA spectral region includes resonance
lines of \ion{Na}{1}, \ion{Al}{1}, \ion{Sc}{2}, \ion{Mn}{2}, \ion{Fe}{1},
and \ion{Co}{1} \citep{morton1991,morton2000}. The stellar spectrum
is dominated by \ion{Mg}{1} and \ion{Fe}{1} lines coming from low energy
levels. Probably, these spectral features, likely to be present also in the
spectrum of the planet atmosphere, produce the observed deeper transit
\citep[see][for a similar case]{vidal04}.

The end of the second exposure is at the phase of the planet ingress, as 
shown in Figure~\ref{lightcurve}. It is notable that the NUVA flux during 
the second exposure lies below the out-of-transit level by $\sim2\sigma$. We 
divided this particular exposure into three equal sub-exposures plotted as 
black crosses. These suggest an early ingress in the NUVA spectral region. 
\subsection{Detection of other elements}
\begin{figure}
\includegraphics[width=13cm]{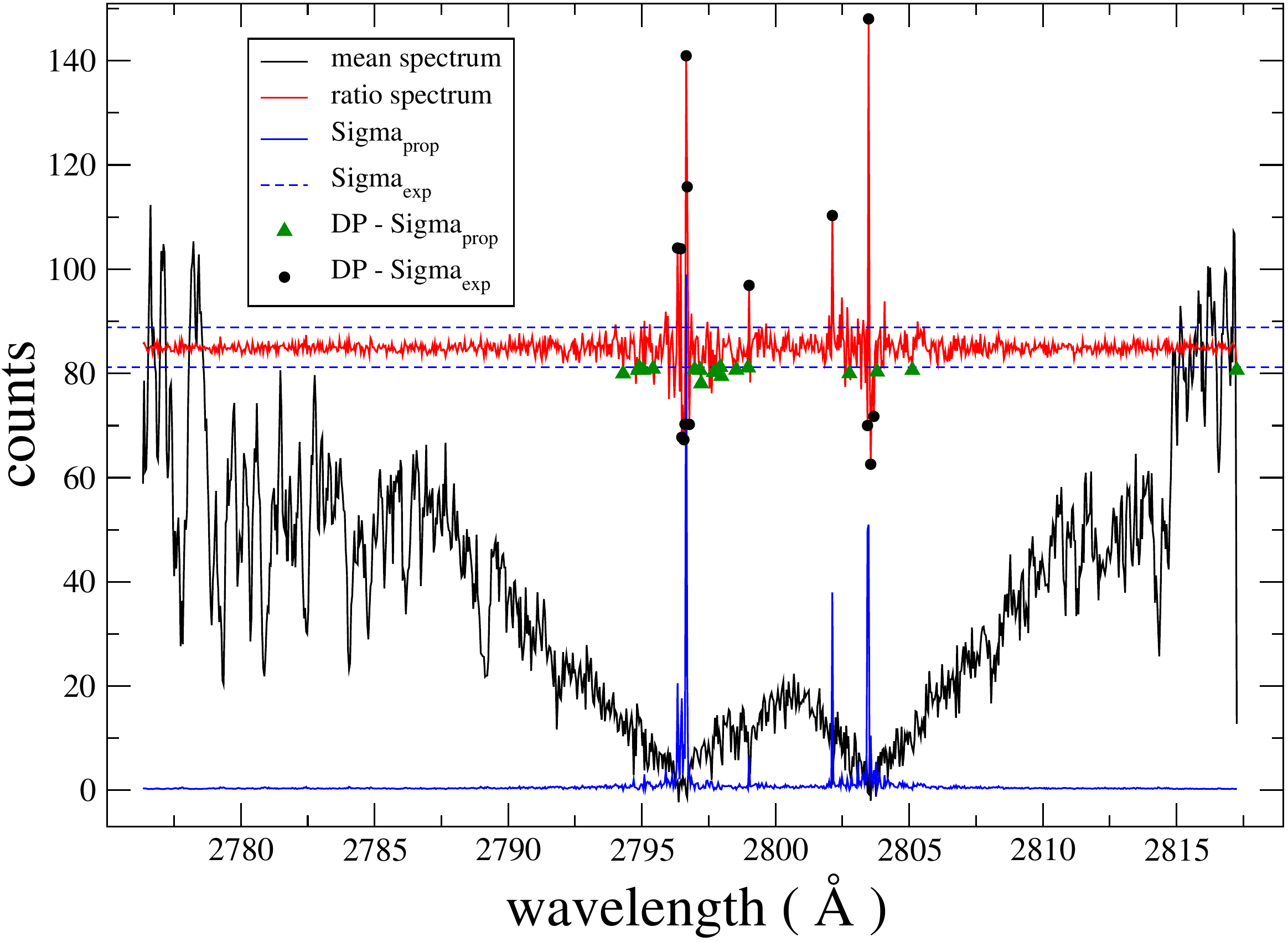}
\caption{\label{Mgline} The black line shows the observed spectrum obtained
averaging the five available COS spectra. The red line shows the $d_{\lambda}$
spectrum, magnified five times and shifted upwards for display reasons.
The blue lines show the $\sigma_{d_{\lambda}|_{prop}}$ spectrum
(full line) and the values of $\sigma_{d_{\lambda}|_{exp}}$ (dashed lines).
The full black circles show the position of the deviating points assuming
$\sigma=\sigma_{d_{\lambda}|_{exp}}$, while the full green triangles show
the position of the deviating points assuming
$\sigma=\sigma_{d_{\lambda}|_{prop}}$.}
\end{figure}

In each of the three observed wavelength ranges we calculated a ratio spectrum
($d_{\lambda}$) between the in-transit spectrum ($in_{\lambda}$) measured in 
the third exposure and the out-of-transit spectrum ($out_{\lambda}$), the 
mean of the first and fifth exposures. To these ratio spectra we associated 
two different uncertainties: (i) the standard deviation from the mean, 
$\bar d$, which we denote $\sigma_{d_{\lambda}|_{exp}}$. (ii) The 
uncertainty for each individual wavelength point in the ratio spectrum from 
the propagated uncertainties. We denote this $\sigma_{d_{\lambda}|_{prop}}$. 
Expressed symbolically:
\begin{equation}
d_{\lambda}=\frac{in_{\lambda}}{out_{\lambda}}
\end{equation}
and
\begin{equation}
\sigma_{d_{\lambda}|_{exp}}=\frac{\sqrt{(\bar{d}-d_{\lambda})^2}}{N-1}  \qquad
\sigma_{d_{\lambda}|_{prop}}=\sqrt{\left(\frac{\sigma_{in_{\lambda}}}{out_{\lambda}}\right)^2+\left(\frac{in_{\lambda}\sigma_{out_{\lambda}}}{out_{\lambda}^2}\right)^2}
\end{equation}
where $N$ is the number of points, $\sigma_{in_{\lambda}}=\sqrt{in_{\lambda}}$,
and $\sigma_{out_{\lambda}}=\sqrt{out_{\lambda}}$. In
NUVA, NUVB, and NUVC $\sigma_{d_{\lambda}|_{exp}}$ is 0.34, 0.12, and 0.76 
respectively. $\sigma_{d_{\lambda}|_{prop}}$ varies with wavelength, as shown in
Fig.~\ref{Mgline}.

Table~\ref{wavelengths} lists the wavelength points of $d_{\lambda}$ (in
laboratory wavelengths) with deviations of more than 3$\sigma$ from $\bar{d}$, 
assuming both $\sigma=\sigma_{d_{\lambda}|_{exp}}$ (left column) and 
$\sigma=\sigma_{d_{\lambda}|_{prop}}$ (right column). Assuming a Gaussian 
distribution and having N=1024$\times$3 wavelength points, we expect 9 points 
in the $d_{\lambda}$ array to fall outside 3$\sigma$ from the mean. Since 
the number of detected deviating wavelength points is much larger than nine 
we looked for correspondences with resonance lines 
\citep{morton1991,morton2000}. Table~\ref{wavelengths} lists the deviating 
wavelength points and the corresponding resonance lines. We include 
occurrences of resonance lines within  a few \kms\ of a deviating wavelength 
point, for example the \ion{Sc}{2} line at 2563.190\,\AA.
\clearpage
\begin{table}
\caption{\label{wavelengths} Wavelength of the spectral points deviating more
than 3$\sigma$, adopting two different $\sigma$s:
$\sigma_{d_{\lambda}|_{exp}}$ and $\sigma_{d_{\lambda}|_{prop}}$. For each
detected deviating point we show the resonance line found lying at the same
position or close to it (*).
In the NUVB region we did not detect any deviating point assuming
$\sigma=\sigma_{d_{\lambda}|_{prop}}$. The deviating
points marked with a {\bf \#} deviate by $\geq$3.5$\sigma$ from the mean.}
\rotatebox{90}{
\tiny{
\begin{tabular}{llll|ll|llll}
\tableline\tableline
Wavelength		       & Resonance & Wavelength 		     & Resonance & Wavelength			  & Resonance & Wavelength		       & Resonance & Wavelength 		     & Resonance \\
$\lambda - 2500$\AA            & line	   & $\lambda - 2500$\AA             & line	 & $\lambda - 2600$\AA            & line      & $\lambda - 2000$\AA            & line	   & $\lambda - 2000$\AA             & line	 \\
3$\times\sigma_{d_{i}|_{exp}}$ & \AA  	   & 3$\times\sigma_{d_{i}|_{prop}}$ & \AA 	 & 3$\times\sigma_{d_{i}|_{exp}}$ & \AA       & 3$\times\sigma_{d_{i}|_{exp}}$ & \AA 	   & 3$\times\sigma_{d_{i}|_{prop}}$ & \AA 	 \\
\multicolumn{4}{c}{NUVA: $\sim$2531 - 2586} & \multicolumn{2}{|c|}{NUVB: $\sim$2650 - 2703} & \multicolumn{4}{c}{NUVC: $\sim$2770 - 2821} \\
\tableline
38.719        & YbII@38.662 &               &             & 63.424        &    	        &		 &  	        & 793.234{\bf \#}& MgII@795.528 \\
38.806        &	            &               &             & 63.549{\bf \#}&     	&		 &  	        & 793.793{\bf \#}& MgII@795.528 \\
39.021        &	            &               &             & 63.674{\bf \#}&     	&		 &  	        & 793.833{\bf \#}& MgII@795.528 \\
              &	            & 40.703{\bf \#}& ScII@40.822 & 66.633{\bf \#}&     	&		 &  	        & 793.993{\bf \#}& MgII@795.528 \\
              &	            & 40.833        & ScII@40.822 & 69.714        & AlII@69.155*&	         &  	        & 794.353{\bf \#}& MgII@795.528 \\
40.876        & ScII@40.822 & 40.876{\bf \#}& ScII@40.822 & 72.459        & ~VII@72.007*& 795.272{\bf \#}& MgII@795.528 &	         &		\\
41.048{\bf \#}&	            &               &             & 72.751{\bf \#}& ~VII@72.007*& 795.391{\bf \#}& MgII@795.528 &	         &		\\
              &	            & 41.135{\bf \#}&             & 78.732{\bf \#}& ~VII@78.575*& 795.431{\bf \#}& MgII@795.528 &	         &		\\
              &	            & 41.178{\bf \#}&             & 83.003{\bf \#}& ~VII@83.090 & 795.511{\bf \#}& MgII@795.528 &	         &		\\
              &	            & 42.126{\bf \#}&             & 89.792        & ~VII@89.884*& 795.551{\bf \#}& MgII@795.528 &	         &		\\
              &	            & 43.893{\bf \#}& ~NaI@43.840 & 97.927{\bf \#}&     	& 795.591{\bf \#}& MgII@795.528 & 	         &		\\
              &	            & 43.893{\bf \#}& ~NaI@43.872 &               &     	& 795.631{\bf \#}& MgII@795.528 & 	         &		\\
              &	            & 45.960{\bf \#}&             &               &     	& 795.711{\bf \#}& MgII@795.528 & 	         &		\\
46.175{\bf \#}& ~SnI@46.548*& 46.175{\bf \#}& ~SnI@46.548*&               &     	&		 &  	        & 795.911{\bf \#}& MgII@795.528 \\
              &	            & 46.691{\bf \#}& ~SnI@46.548*&               &     	&		 &  	        & 796.110{\bf \#}& MgII@795.528 \\
              &	            & 48.369{\bf \#}&EuIII@48.583*&               &     	&		 &  	        & 796.150        & MgII@795.528 \\
              &	            & 49.402{\bf \#}&             &               &     	&		 &  	        & 796.629{\bf \#}& MgII@795.528 \\
              &	            & 49.445{\bf \#}&             &               &     	&		 &  	        & 796.869        & MgII@795.528 \\
              &	            & 50.004{\bf \#}&             &               &     	&		 &  	        & 796.909{\bf \#}& MgII@795.528 \\
62.363{\bf \#}&	            &               &             &               &     	&		 &  	        & 797.468{\bf \#}& MgII@795.528 \\
              &	            & 62.535{\bf \#}&             &               &     	&		 &  	        & 797.907{\bf \#}& MgII@795.528 \\
              &	            & 62.620{\bf \#}&             &               &     	& 797.947	 & ~MnI@798.269*& 	         &		 \\
              &	            & 63.348{\bf \#}& ScII@63.190*&               &     	& 801.059{\bf \#}& ~MnI@801.082 & 	         &		\\
63.391{\bf \#}& ScII@63.190*&               &             &               &     	&		 &  	        & 801.697{\bf \#}& MgII@802.705 \\
63.477        & ScII@63.190*&               &             &               &     	& 802.375{\bf \#}& MgII@802.705 & 	         &		\\
              &	            & 63.563        & ScII@63.190*&               &     	& 802.415{\bf \#}& MgII@802.705 & 	         &		\\
75.999{\bf \#}& MnII@76.106 &               &             &               &     	& 802.495{\bf \#}& MgII@802.705 & 	         &		\\
76.085{\bf \#}& MnII@76.106 &               &             &               &     	& 802.614	 & MgII@802.705 & 	         &		 \\
76.127{\bf \#}& MnII@76.106 &               &             &               &     	&		 &  	        & 802.734{\bf \#}& MgII@802.705 \\
              &	            &               &             &               &     	&		 &  	        & 804.049{\bf \#}& MgII@802.705 \\
              &	            &               &             &               &     	&		 &  	        & 816.182{\bf \#}&		 \\
\tableline
\end{tabular}
}}
\end{table}

In the NUVA wavelength region and adopting $\sigma=\sigma_{d_{\lambda}|_{exp}}$
we obtained 3$\sigma$ deviations corresponding to the position of three
resonance lines: \ion{Yb}{2} at 2538.662\,\AA, \ion{Sc}{2} at 2540.822\,\AA, 
and \ion{Mn}{2} at 2576.106\,\AA. Assuming instead 
$\sigma=\sigma_{d_{\lambda}|_{prop}}$ \ion{Sc}{2} at 2540.822\,\AA\ and 
the \ion{Na}{2} doublet at 2543.8\,\AA\ are picked out. In the NUVB 
region we only find the \ion{V}{2} line at 2683.090\,\AA\ and only assuming 
$\sigma=\sigma_{d_{\lambda}|_{exp}}$. However three other \ion{V}{2} 
resonance lines and an \ion{Al}{2} line lie close to other detected 
deviating points. In the NUVC region we recognize immediately that most 
of the deviating points are in the core of the \ion{Mg}{2} resonance
lines, both assuming $\sigma=\sigma_{d_{\lambda}|_{exp}}$ and
$\sigma=\sigma_{d_{\lambda}|_{prop}}$. We also pick out
the \ion{Mn}{1} line at 2801.082\,\AA, while the \ion{Mn}{1} line at
2798.269\,\AA\ lies close to the wavelength of another deviating point.

Figure~\ref{Mgline} shows the cores of the
\ion{Mg}{2} resonance lines. We show the observed spectrum, $d_{\lambda}$,
$\sigma_{d_{\lambda}|_{prop}}$, and the deviating wavelength points both
assuming $\sigma=\sigma_{d_{\lambda}|_{exp}}$ and
$\sigma=\sigma_{d_{\lambda}|_{prop}}$. With
$\sigma=\sigma_{d_{\lambda}|_{exp}}$ the deviating points 
correspond to the core of the \ion{Mg}{2} line where the signal level is
low. This is to be expected: the low count rates at these wavelengths lead 
$d_{\lambda}$ to be very noisy here.
In contrast,
with $\sigma=\sigma_{d_{\lambda}|_{prop}}$
each element of the $d_{\lambda}$ spectrum is assessed against its own 
Poisson error. In this case the deviating points are all below the mean rate 
spectrum, and the deviating points appear at the margins of the line core. 
These points indicate excess \ion{Mg}{2} absorption during transit. This is 
attributable to absorption by the planet's atmosphere. This pattern is seen 
not only for the two \ion{Mg}{2} resonance lines, but also for the \ion{Sc}{2} 
line at 2563.190\,\AA. This line, together with the \ion{Mn}{1} line at 
2798.269\,\AA, has the intriguing property that the difference between the 
position of the resonance line and of the detected deviating wavelength 
point(s) corresponds to a velocity of $\sim$30\,\kms\  (about 3 resolution 
elements), close to the planet escape velocity of 
$\sim$37\,\kms\ \citep{hebb2009},
although it would not then be clear why this pattern does
not appear also for other detected lines of the same ion.
\clearpage

There are 95 known resonance lines lying within the observed wavelength 
ranges, including those of heavy elements. This is a small minority of the 
more than 4000 lines present in the stellar spectrum. The fact that we find 
deviating points predominantly at wavelengths corresponding to resonance 
lines strongly suggests  we are detecting features produced by the planet 
atmosphere. Reassuringly, with either definition of $\sigma$, we obtained 
nine points that do not match any known resonance line, in perfect accordance 
with statistical expectations. We repeated the exercise picking out 
deviations in excess of 3.5$\sigma$, obtaining almost the same deviating 
points at the position of known resonance lines and fewer points where no 
resonance lines were found.
\section{Discussion}\label{sec:disc}
We have performed three independent analyses, each of which suggests absorption
in the resonance lines of metals from an extended atmosphere surrounding
the transiting planet WASP-12b. In Section~\ref{charb} we found a deeper 
transit  in the core region of the \ion{Mg}{2} doublet at the $2.8\sigma$ level.

In Sect.~\ref{sec_lightcurve}, the transit depths in the NUVA, NUVB, and 
NUVC wavelength ranges respectively imply effective planet radii of 
2.69$\pm$0.24\,R$_J$, 2.18$\pm$0.18\,R$_J$, and 2.66$\pm$0.22\,R$_J$.
WASP-12b's optical radius is $R_{\rm P} = 1.79 \pm 0.09 \,R_J$ while the 
mean Roche lobe radius is 2.36\,R$_J$ using Paczy{\'n}ski's (1971) 
prescription.

Table~\ref{wavelengths} shows that we detect enhanced transit depths
at the wavelengths of resonance lines of neutral sodium, tin and manganese,
and at singly ionised ytterbium, scandium, manganese, aluminum, vanadium and 
magnesium. Finally we detect an enhanced transit depth within $0.12$\AA\ of a 
resonance line of doubly ionised europium. We also find the statistically 
expected number of anomalous transit depths at wavelengths not associated 
with any known resonance line.

Taken as a whole, these results constitute compelling 
evidence that WASP-12b is surrounded by an exosphere which over-fills the 
planet's Roche lobe, confirming predictions by \citet{li2010}. This 
exosphere is likely composed of a number of elements/ions, including probably 
\ion{Na}{1}, \ion{Mg}{1}, \ion{Mg}{2}, \ion{Al}{1}, \ion{Sc}{2}, 
\ion{Mn}{2}, \ion{Fe}{1}, and \ion{Co}{1}.
The phenomenon found in HD209458b \citep{vidal03,vidal08} probably occurs 
generally for hot Jupiter exoplanets. By analogy with HD209458b, and 
as WASP-12b and its host star are almost certainly predominantly composed 
of hydrogen, we expect that this exosphere is hydrogen rich.

Models by \citet{yelle} suggest that elements other then H and He should
not be present in the upper atmosphere due to the low vertical mixing rate,
but this takes Jupiter as the starting point. WASP-12b is
extremely close to the host star and consequently the stellar irradiation and
tidal effects could induce prodigious mixing, affecting the chemistry of 
the planet atmosphere. 
Our detections of several metallic elements and/or ions is certainly 
consistent with a metal-rich atmosphere for WASP-12b.

The most surprising result is provided by the juxtaposition of our data with 
the optical ephemeris. We took contemporaneous optical photometry with 
OU-OAM PIRATE \citep{klb2009} which showed the ephemeris of \citet{hebb2009} 
remains accurate. Figure~\ref{lightcurve} shows the NUVA transit has an early 
ingress and an egress consistent with the optical ephemeris. In contrast, naive 
momentum considerations and hydrodynamic simulations would instead suggest 
that the effect of a diffuse cloud surrounding the planet would be to smear 
and delay egress while ingress is relatively unaffected, see e.g. 
Fig.~1 and 2 of \citet{schneiter2007}.

In detail, the shape of the diffuse cloud may well be
element/ion dependent since different elements/ions behave
differently in the presence of strong radiation pressure. This can explain
why we observe different transit shapes in the NUVA region and the other 
regions. As Fig.~\ref{spectra} shows, the stellar spectrum in the NUVA 
region is strongly absorbed by a plethora of lines, dominated those of
neutral elements.  The NUVC region is also strongly absorbed in the stellar 
photosphere but predominantly from the \ion{Mg}{2} doublet. It is presumably 
the cumulative absorption from many relatively weak spectral lines in the 
planet's exosphere which creates the excess transit depth in the NUVA region, 
while Table~\ref{wavelengths} and Eq.~\ref{mg2depth} demonstrate that 
planet's absorption in the NUVC region is associated with the \ion{Mg}{2} 
doublet. The \ion{Mg}{2} ion will experience different forces to neutral 
atoms in an environment where there is certainly a strong radiation field, 
and  strong and varying large-scale magnetic fields are also likely. The NUVB 
light curve is least deviant from the optical transit, and this is 
consistent with the relative dearth of strongly absorbing lines in this 
spectral window, c.f. Fig.~\ref{spectra}.

We do not have any detailed explanation for the observed early ingress in 
NUVA, but we speculate the effect could be produced  if material is lost 
from the planet exosphere and forms a diffuse ring or torus around the 
star enveloping the planet's orbital path, as models suggest \citep{li2010}. 
The orbital motion of the planet through this medium might compress the 
material in front of it. This could increase the opacity of the medium 
through which the star is viewed immediately before first contact. A void 
in the medium might be expected to form behind the planet, and consequently 
the egress is relatively unaffected by the diffuse ring.

Our observations demonstrate that COS spectroscopy of transiting exoplanets 
has the potential to detect many species via transmission spectroscopy, 
and to measure velocities and deduce spatial distributions. There are now 
about 40 known transiting exoplanets with orbital periods shorter than that 
of HD209458b. Many of these transit stars significantly brighter than 
WASP-12b. COS spectroscopy of brighter examples will allow us probe the 
exosphere species-by-species examining their density, velocity and spatial 
distributions. This detailed information should allow us to determine whether 
these planets really are being photo-evaporated by their host stars, and, 
if so, to empirically deduce the mass loss rate. We encourage detailed 
element/ion dependent modeling of the exosphere in the highly irradiated 
environment of  WASP-12b and similar systems, and observations of other 
similar extrasolar planets. There is a rich new parameter space to explore!
\acknowledgments
Astronomy research at the Open University is supported by an STFC rolling grant.
We thank O. Kochukhov, D. Shulyak, and T. Ryabchikova for the useful 
discussions. LF thanks the whole CASA staff for the hospitality and the 
fruitful discussions, particularly Steven Penton, St\'{e}phane B\'{e}land, 
Kevin France, Tom Ayres and Eric Burgh. Support for program \#11651 was 
provided by NASA through a grant from the Space Telescope Science Institute, 
which is operated by the Association of Universities for Research in 
Astronomy, Inc., under NASA contract NAS 5-26555.

{\it Facilities:} \facility{HST (COS)}.

\end{document}